\documentclass[preprint2]{aastex62}

\newcommand\aastex{AAS\TeX}

\usepackage{epstopdf}

\received{November 1, 2018}
\revised{November 1, 2018}
\accepted{\today}

\submitjournal{ApJ}

\shorttitle{\aastex\ Chrmospheric evaporation in a circular-ribbon flare}
\shortauthors{Zhang et al.}

\begin{document}

\title{Imaging observations of chromospheric evaporation in a circular-ribbon flare}

\correspondingauthor{Q. M. Zhang}
\email{zhangqm@pmo.ac.cn}

\author[0000-0003-4078-2265]{Q. M. Zhang}
\affil{Key Laboratory for Dark Matter and Space Science, Purple Mountain Observatory, CAS, Nanjing 210034, China}

\author[0000-0002-4538-9350]{D. Li}
\affil{Key Laboratory for Dark Matter and Space Science, Purple Mountain Observatory, CAS, Nanjing 210034, China}
\affil{State Key Laboratory of Space Weather, Chinese Academy of Sciences, Beijing 100190, PR China}
\affil{CAS Key Laboratory of Solar Activity, National Astronomical Observatories, Beijing 100012, China}

\author{Y. Huang}
\affil{Key Laboratory for Dark Matter and Space Science, Purple Mountain Observatory, CAS, Nanjing 210034, China}

\begin{abstract}
In this paper, we report our multiwavelength imaging observations of chromospheric evaporation in a C5.5 circular-ribbon flare (CRF) on 2014 August 24.
The flare was observed by the Atmospheric Imaging Assembly (AIA) 
on board the \textit{Solar Dynamics Observatory} (\textit{SDO}), X-ray Telescope (XRT) on board the \textit{Hinode} spacecraft, and ground-based Nobeyama Radioheliograph (NoRH).
The CRF consisted of a discrete circular ribbon with a diameter of $\sim$1$\arcmin$ and a short inner ribbon observed in ultraviolet (UV), extreme-ultraviolet (EUV), soft X-ray (SXR), 
and especially in 17 GHz. The peak time ($\sim$04:58 UT) of the flare in 17 GHz coincided with that in UV 1600 {\AA} and SXR derivative as a hard X-ray proxy, implying the peak time of 
impulsive energy deposition in the lower atmosphere.
Shortly after the peak time, converging motion and filling process in the flare loop were revealed in AIA 131 {\AA} and two XRT filters (Be\_thin and Be\_med), which are clear evidence for 
chromospheric evaporation upflows. The chromospheric evaporation lasted for $\sim$6 minutes until $\sim$05:04 UT. The temperature, density, and apparent velocities of the upflows are 
$\sim$10$^7$ K, $\sim$1.8$\times$10$^{10}$ cm$^{-3}$, and 50$-$630 km s$^{-1}$ with a mean value of $\sim$170 km s$^{-1}$. 
By comparison with previous models, we are able to estimate that energies above 5$\times$10$^{10}$ erg cm$^{-2}$ s$^{-1}$ are likely needed to explain the observational results.
Since heating by thermal conduction does not seem to provide enough energy, alternative mechanisms such as nonthermal electrons or Alfv\'{e}nic waves might need to be invoked.
\end{abstract}

\keywords{Sun: chromosphere --- Sun: flares --- Sun: UV radiation --- Sun: X-rays, gamma rays}

\section{Introduction} \label{sec:intro}
Solar flares are impulsive increases of electromagnetic emissions from radio wave to $\gamma$-ray as a result of the release of magnetic free energy of 
10$^{29}$$-$10$^{32}$ erg \citep{fle11,shi11,war16a,war16b}. The accumulated magnetic energy is converted to the kinetic and thermal energy of the reconnection outflows 
as well as the nonthermal energy of accelerated particles via magnetic reconnection \citep[e.g.,][]{hol11,tian14,yang15}. In the two-dimensional (2D) standard flare model, 
the high-energy electrons propagate downward along the reconnected magnetic field and precipitate in the chromosphere \citep{car64,stu66,hir74,kop76}. 
The collision of electrons with ions result in impulsive heating of the localized plasma and increases of radiations in hard X-ray (HXR) and microwave \citep{bro71,bas98}. 
If the heating rate in the chromosphere is significantly larger than the energy dissipation rate, the hot plasmas would flow upward along the flare loops driven by overpressure, 
a process called chromospheric evaporation \citep{act82,fis85a,fis85b,fis85c,can90,abb99,all05}. Meanwhile, the dense plasmas experience downward motion 
at much smaller speeds due to the momentum balance, a process called chromospheric condensation \citep{fis89,wue94}. There are two types of chromospheric evaporation. 
For the explosive evaporation, the emission lines formed in the corona are blueshifted, while the emission lines formed in the transition region and chromosphere 
are redshifted \citep[e.g.,][]{cza99,bro04,mil09,bro13,li15}. The velocities of the blushifted upflows are 100$-$800 km s$^{-1}$, 
while the velocities of the redshifted downflows are tens of km s$^{-1}$. For the gentle evaporation, the lines can show blue- or no shift \citep[e.g.,][]{mil06,bat09,sad15}. 
Based on the assumption that the electron beams last for 5 s, with a fixed energy flux, a fixed spectral index ($\delta=4$), and a fixed low-energy cut-off ($E_c=20$ keV), 
\citet{fis85c} derived the threshold for the input energy flux of explosive evaporation ($\sim$10$^{10}$ erg cm$^{-2}$ s$^{-1}$). However, it is recently found that the threshold 
($F$) depends strongly on the electron energy ($E_\ast$) and duration of heating \citep{reep15}. The relationship between $F$ and $E_\ast$ is linearly fitted in log-log space,
$\log_{10}F=6.99+2.43\log_{10}E_\ast$ (see their Fig. 7).
Chromospheric evaporations mainly take place in the impulsive and decay phases of flares. However, they occasionally occur in the pre-flare phase \citep{bro10,li18}.
For the driving mechanisms of chromospheric evaporation, the role of nonthermal electrons has been richly investigated and widely accepted \citep{reep15,rub15}. 
In some cases, thermal conduction plays an essential role \citep{fal97,bat09}.

There are abundant observations of chromospheric evaporation in solar flares, most of which are spectroscopic \citep[e.g.,][]{you13,you15,pol15,pol16,mil15,tian18}. So far, direct imaging 
observations of chromospheric evaporation are rare owing to the limited spatial and temporal resolutions of solar telescopes. One way of detecting the evaporation upflow is the rapid lifting 
and converging motions of double sources along the flare loops in HXR and microwave wavelengths \citep{asch95,liu06,ning09,ning10}. The other way is direct imaging of the upflow in 
extreme-ultraviolet (EUV) and SXR wavelengths \citep[e.g.,][]{sil97,nit12,zqm13,li17a}.

Circular-ribbon flares (CRFs) are observed and investigated in detail by \citet{mas09}. As its name implies, the ribbons of CRFs have circular or quasi-circular shapes 
in Ca {\sc ii} H, H$\alpha$, UV, and EUV wavelengths \citep{sun12,wang12,jia13,kum16,hao17,li17b,song18}. 
The magnetic topology of CRFs are mostly associated with a magnetic null point ($\mathbf{B}=0$), a spine, and a dome-like fan surface \citep{zqm12,zqm15}. 
\citet{zqm16a} studied a C4.2 CRF on 2015 October 16 observed by the \textit{Interface Region Imaging Spectrograph} \citep[\textit{IRIS;}][]{dep14}. For the first time, 
the authors found explosive chromospheric evaporation in the circular ribbon (CR) and inner ribbon (IR). Upflows at a speed of 35$-$120 km s$^{-1}$ are observed in the high-temperature
Fe {\sc xxi} $\lambda$1354.09 line ($\log T\approx7.05$), and downflows at a speed of 10$-$60 km s$^{-1}$ are observed in the low-temperature Si {\sc iv} $\lambda$1393.77 line ($\log T\approx4.8$).
In a follow-up work, \citet{zqm16b} reported periodic chromospheric condensation in a homologous CRF in the same active region (AR).
However, direct imaging observation of chromospheric evaporation in CRFs has never been investigated. In this paper, we report our multiwavelength observations of chromospheric evaporation 
in a C5.5 CRF on 2014 August 24. In Section~\ref{s:data}, we describe the observations and data analysis. Results are present in Section~\ref{s:res}. We compare our findings with previous works
in Section~\ref{s:disc} and give a brief summary in Section~\ref{s:sum}.

\section{Data analysis} \label{s:data}
The C5.5 flare took place in NOAA AR 12149 (N10E44). It was observed by the the Atmospheric Imaging Assembly \citep[AIA;][]{lem12} on board the \textit{Solar Dynamics Observatory} (\textit{SDO}) 
spacecraft. AIA takes full-disk images in two UV (1600 and 1700 {\AA}) and seven EUV (94, 131, 171, 193, 211, 304, and 335 {\AA}) wavelengths. The photospheric line-of-sight (LOS) magnetograms 
were observed by the Helioseismic and Magnetic Imager \citep[HMI;][]{sch12} on board \textit{SDO}. The AIA and HMI level\_1 data were calibrated using the standard \textit{Solar Software} 
(\textit{SSW}) program \textit{aia\_prep.pro} and \textit{hmi\_prep.pro}, respectively. The flare was also captured by the X-ray Telescope \citep[XRT;][]{gol07} on board the \textit{Hinode} \citep{kos07} 
spacecraft with a smaller field of view (384$\arcsec$$\times$384$\arcsec$). The SXR images observed by the Be\_thin and Be\_med filters were calibrated using the standard 
\textit{SSW} program \textit{xrt\_prep.pro} and coaligned with the AIA 131 {\AA} images. SXR fluxes of the flare in 0.5$-$4 {\AA} and 1$-$8 {\AA} were recorded by the \textit{GOES} spacecraft. 
The Nobeyama Radioheliograph \citep[NoRH;][]{naka94} at the Nobeyama Radio Observatory also observed this flare. As a ground-based radio telescope, NoRH observes the full disk
at frequencies of 17 and 34 GHz with spatial resolutions of 10$\arcsec$ and 5$\arcsec$, respectively. 
The observational parameters, including the instrument, wavelength, time cadence, and pixel size are summarized in Table~\ref{tab:para}.

\begin{deluxetable}{ccccc}
\tablecaption{Description of the observational parameters \label{tab:para}}
\tablecolumns{5}
\tablenum{1}
\tablewidth{0pt}
\tablehead{
\colhead{Instru.} &
\colhead{$\lambda$} &
\colhead{Time} & 
\colhead{Cad.} & 
\colhead{Pix. size} \\
\colhead{} & 
\colhead{({\AA})} &
\colhead{(UT)} & 
\colhead{(s)} & 
\colhead{(\arcsec)}
}
\startdata
\textit{SDO}/AIA & 94$-$335 & 04:30$-$06:00 & 12 & 0.6 \\
\textit{SDO}/AIA & 1600 & 04:30$-$06:00 & 24 & 0.6 \\
\textit{SDO}/HMI & 6173 & 04:30$-$06:00 & 45 & 0.6 \\
\textit{Hinode}/XRT & Be\_thin & 04:33$-$05:41 & $\sim$30 & 1.03 \\
\textit{Hinode}/XRT & Be\_med & 04:58$-$05:08 & $\sim$20 & 1.03 \\
\textit{GOES}     & 0.5$-$4 & 04:30$-$06:00  & 2.05 & \nodata \\
\textit{GOES}     & 1$-$8    & 04:30$-$06:00  & 2.05 & \nodata \\
NoRH                 & 17 GHz & 04:30$-$06:00  & 1 & 5 \\
\enddata
\end{deluxetable}

In order to have a better evaluation of the temperature and density of the flare, we performed differential emission measure (DEM) analysis using the simultaneous AIA images in six EUV 
wavelengths (94, 131, 171, 193, 211, 335 {\AA}). The EUV flux of the optically thin plasma at a certain passband is expressed as 
\begin{equation} \label{eqn-1}
F_{i}=\int_{T_1}^{T_2} R_{i}(T)\times \mathrm{DEM}(T)dT,
\end{equation}
where $R_{i}(T)$ is the temperature response function of passband $i$, $T_1$ and $T_2$ represent the minimum and maximum temperatures, and $\mathrm{DEM}(T)$ stands for DEM 
as a function of $T$ \citep{zqm14,zqm16c}. 
The total column emission measure ($\mathrm{EM}$) along the LOS depth ($H$) is defined as the integral of $\mathrm{DEM}(T)$, 
\begin{equation}  \label{eqn-2}
\mathrm{EM}=\int_{T_1}^{T_2}\mathrm{DEM}(T)dT\approx n_{\mathrm e}^2H,
\end{equation}
where $n_{\mathrm e}$ denotes the electron number density. The DEM-weighted average temperature ($\bar{T}$) is expressed as
\begin{equation} \label{eqn-3}
\bar{T}=\frac{\int_{T_1}^{T_2}\mathrm{DEM}(T)\times TdT}{\mathrm{EM}}.
\end{equation}
To improve the signal to noise ratio, we performed a 2$\times$2 binning of the images. A small patch of quiet region outside the flare region is taken as the background, whose intensities 
are removed before conducting the DEM analysis. Besides, we take $\log T_1=5.5$ and $\log T_2=7.5$ in the inversion. The method and code are the same as those we previously 
used \citep{zqm14,zqm16c}.

\section{Results} \label{s:res}
\subsection{Circular-ribbon flare and jets} \label{s:flare}
\begin{figure*}
\plotone{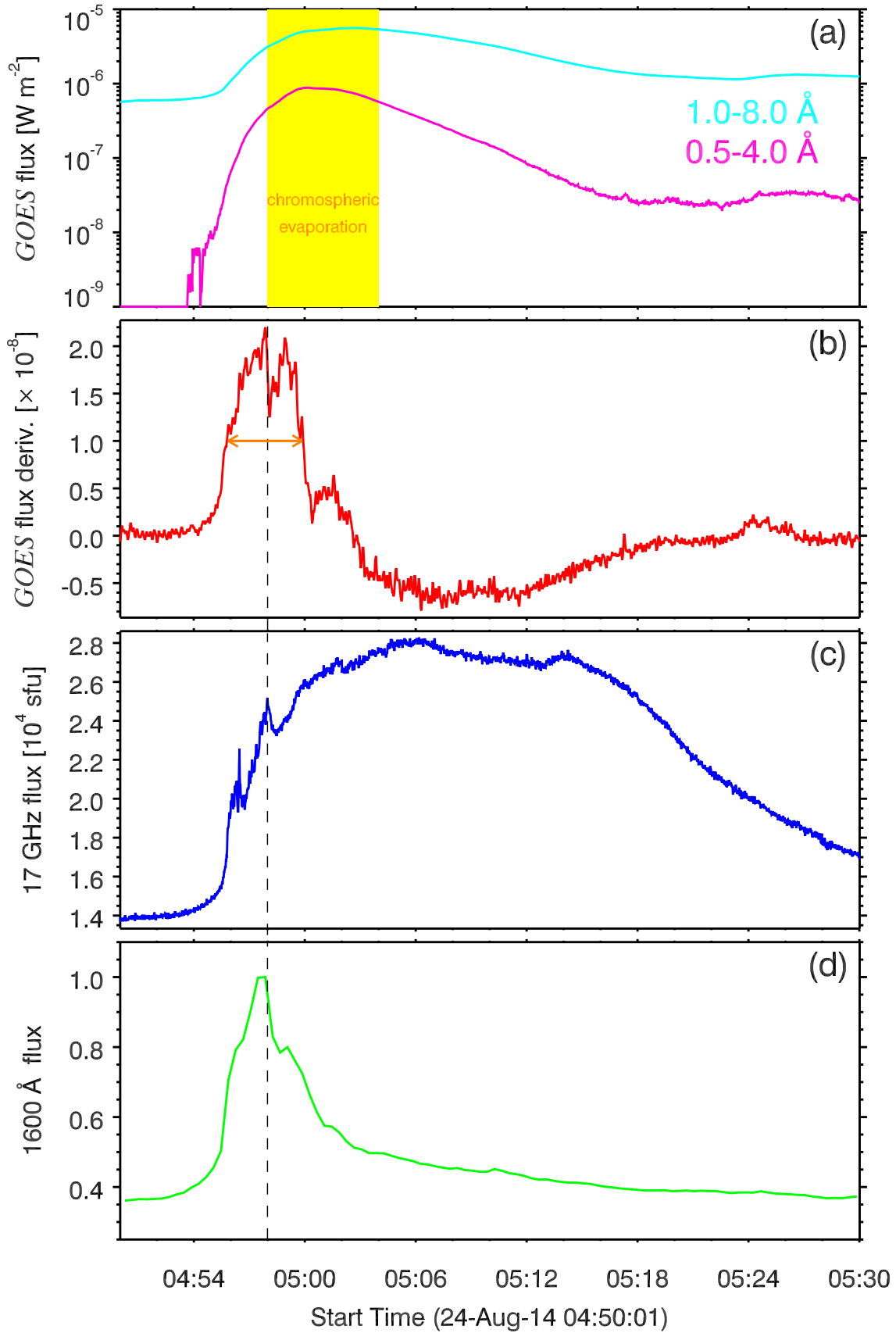}
\caption{(a)-(d) Light curves of the C5.5 flare in SXR, HXR proxy (time derivative of the light curve in 1$-$8 {\AA}), 17 GHz, and 1600 {\AA} (normalized).
             The black dashed line denotes the time at 04:57:58 UT. In panel (a), the yellow region stands for the time of chromospheric evaporation.
             In panel (b), the orange horizontal arrow denotes the duration of heating by nonthermal electrons ($\sim$4 minutes).
\label{fig1}}
\end{figure*}

In Figure~\ref{fig1}(a), the SXR light curves of the flare in 0.5$-$4 {\AA} and 1$-$8 {\AA} are plotted with magenta and cyan lines, respectively. The fluxes increase slowly during the 
pre-flare phase (04:50$-$04:55 UT). Afterwards, the fluxes increase rapidly during the impulsive phase (04:55$-$05:02 UT). Then, the emissions decline gradually until $\sim$05:25 UT. 
Hence, the lifetime of the flare is about 0.5 hr. Considering that HXR observation of the flare during the impulsive phase is unavailable, we take the time derivative of the 
light curve in 1$-$8 {\AA} as a HXR proxy based on the Neupert effect \citep{neu68}, which is plotted in Figure~\ref{fig1}(b). Two peaks at 04:58 UT and 04:59 UT are noticeable. 
The duration of chromospheric heating by nonthermal electrons is defined as the full width at half maximum (FWHM) of the HXR light curve, which is indicated by the orange horizontal arrow.
In order to derive the light curve in 17 GHz, we integrate the intensities of the whole flare region (see Figure~\ref{fig3}(b)). The light curve is plotted in Figure~\ref{fig1}(c). 
Two sharp spikes at 04:56:27 UT and 04:57:58 UT superposed on the gradual component are identified. The second spike coincides with the first peak in HXR proxy indicated by the 
black dashed line. The spikes in microwave are in fact clear evidence of gyrosynchroton emission by nonthermal electrons.
Likewise, we derived the UV light curve by integrating the intensities of the flare region in 1600 {\AA} (see Figure~\ref{fig3}(a)). The normalized curve is 
plotted in Figure~\ref{fig1}(d). The same peak shows up at 04:58 UT, although the time cadence is relatively lower. The weaker peak around 04:59 UT is coincident with that in HXR proxy.
Combining the light curves in multiwavelengths, we conclude that the peak of the energy deposition in the lower atmosphere carried by nonthermal electrons took place around 04:58 UT.

\begin{figure*}
\plotone{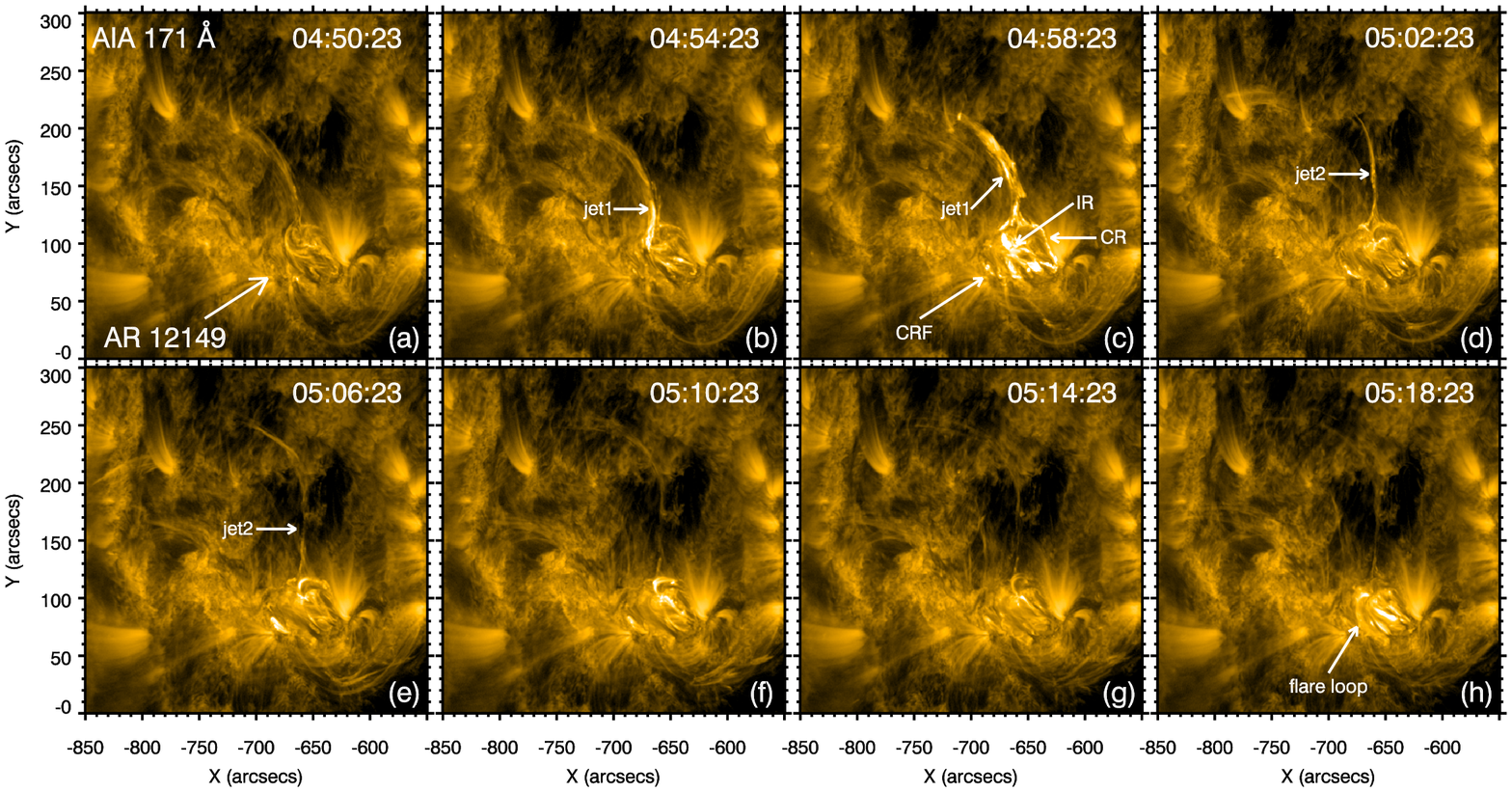}
\caption{Eight snapshots of the AIA 171 {\AA} images during 04:50$-$05:18 UT.
            The white arrows point to AR 12149, CRF, CR, IR, jet1, and jet2.
\label{fig2}}
\end{figure*}

\begin{figure*}
\plotone{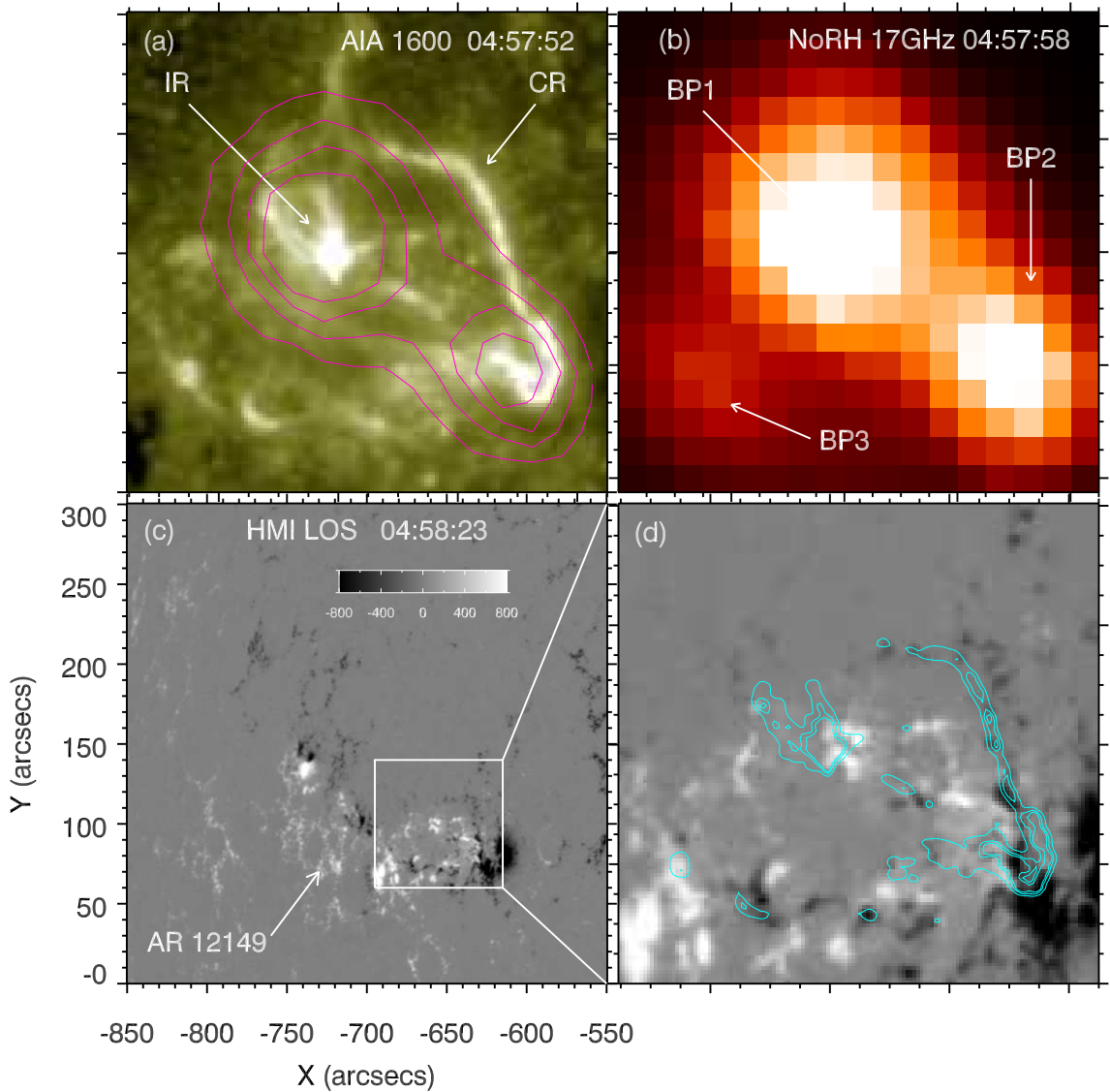}
\caption{\textit{Bottom panels:} HMI LOS magnetogram at 04:58:23 UT and a closeup of the flaring region.
              White and black colors represent positive and negative polarities.
              \textit{Top panels:} AIA 1600 {\AA} image and NoRH 17 GHz image around 04:58 UT with the same FOV as panel (d).
              Intensity contours (50\%, 70\%, 90\%) of the radio image are superposed on panel (a) with orange lines.
              Intensity contours of the 1600 {\AA} image are superposed on panel (d) with cyan lines. 
\label{fig3}}
\end{figure*}

In Figure~\ref{fig2}, the whole evolution of the flare is represented by eight snapshots in 171 {\AA}. In the pre-flare phase, AR 12149 was somewhat quiet (see panel (a)). 
Four minutes later, the first jet (jet1) appeared and propagated in the northeast direction along a closed coronal loop (see panel (b)). At $\sim$04:58 UT, the intensities of the 
discrete CR and IR of the CRF reached their maxima (see panel (c)). Soon after, a second jet (jet2) spurted out of the flare in the north direction before deflecting eastward 
(see panels (d)-(e)). The intensities of flare loops and jets decreased gradually and faded out (see panel (h)).

In Figure~\ref{fig3}(c), the HMI LOS magnetogram at 04:58:23 UT is displayed in grayscale. The white arrow points to AR 12149 with mixed polarities. A closeup of the flare region
with a field of view (FOV) of 80$\arcsec$$\times$80$\arcsec$ is demonstrated in panel (d). The most conspicuous feature of the magnetogram is that positive polarities are surrounded by 
negative polarities. The AIA 1600 {\AA} image and NoRH 17 GHz image around 04:58 UT with the same small FOV are displayed in the top panels. We superpose the intensity contours of
the UV image on panel (d) with cyan lines. It is obvious that the IR is cospatial with the positive polarities and the discrete CR is cospatial with negative polarities, which is similar to the case 
of C-class CRFs on 2015 October 16 \citep{zqm16a,zqm16b}. Such a correspondence between the flare ribbons and magnetic polarities is strongly suggestive of the existence of a magnetic
null point and the fan-spine configuration in the corona. The radio image, although with a lower resolution, features three bright patches (BPs) pointed by the white arrows. 
Likewise, we superpose the intensity contours of radio image on panel (a) with orange lines. The most prominent patch (BP1) is cospatial with the IR and the weaker surrounding patches 
(BP2 and BP3) are cospatial with the CR. To our knowledge, this is the first detection of distinguishable ribbons of CRFs in microwave.

\begin{figure*}
\plotone{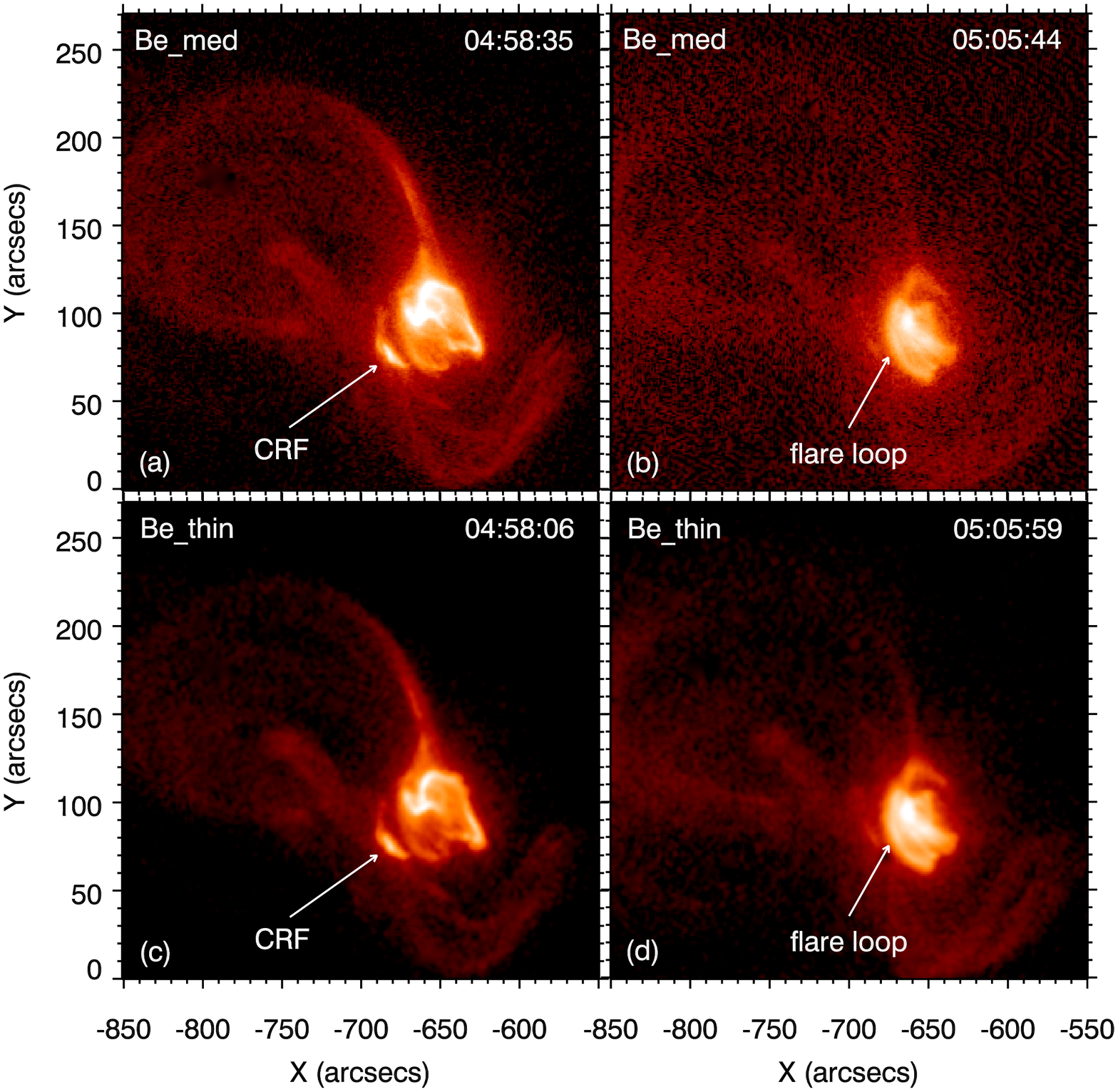}
\caption{SXR images taken by XRT/Be\_med and XRT/Be\_thin filters around 04:58 UT (\textit{left panels}) and 05:06 UT (\textit{right panels}).
            The white arrows point to the CRF and flare loop.
\label{fig4}}
\end{figure*}

In Figure~\ref{fig4}, the left panels show images taken by the XRT filters around 04:58 UT. The SXR emissions of the flare come from the hot plasmas of several MK and have similar morphology 
to the EUV images (see Figure~\ref{fig2}(c)). The large-scale coronal loop above the flare is the loop that guides jet1.

\subsection{Chromospheric evaporation} \label{s:che}
\begin{figure*}
\plotone{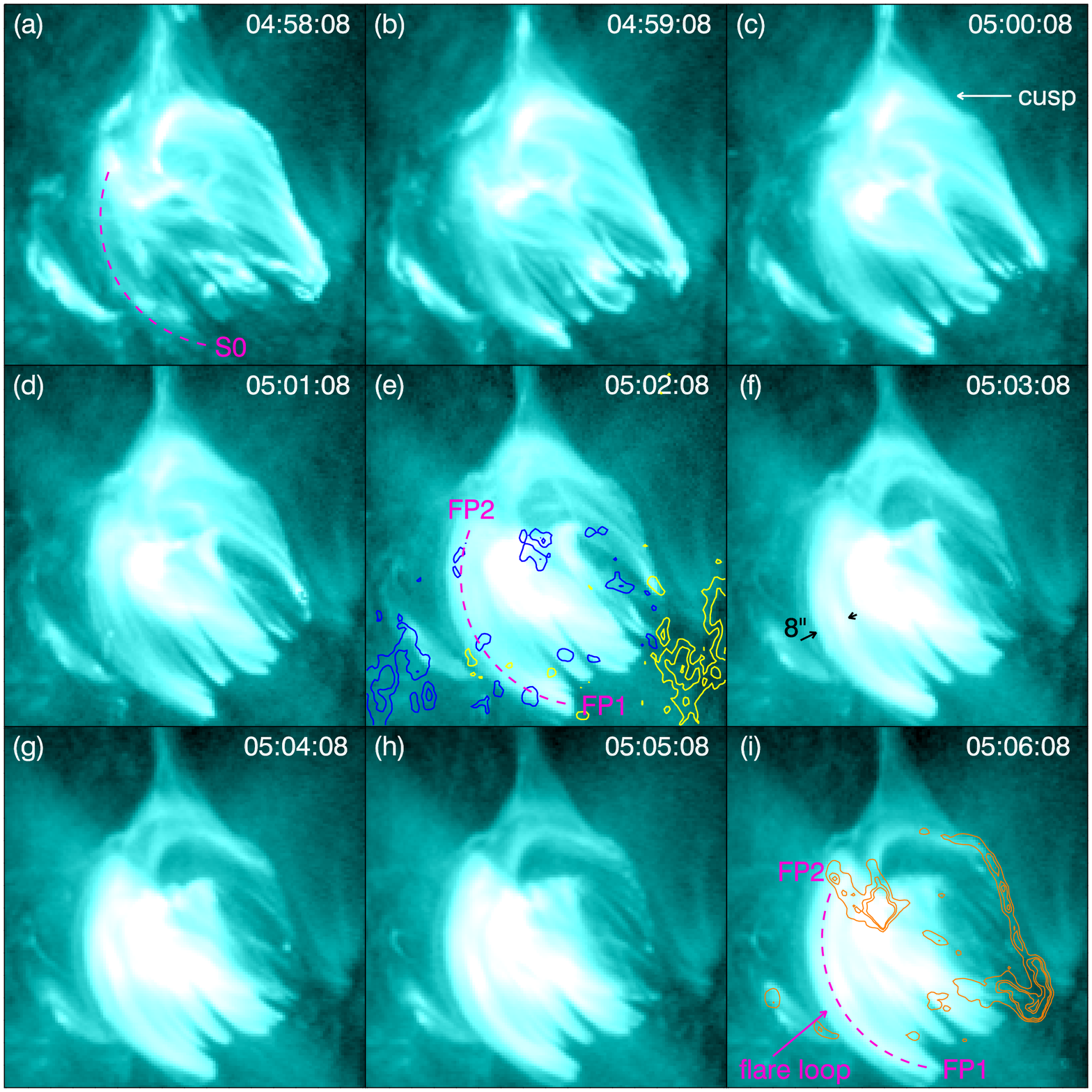}
\caption{AIA 131 {\AA} images during the chromospheric evaporation.
             In panels (a), (e), and (i), the short slice (S0) along the hot flare loop is labeled with a magenta dashed line.
             ``FP1" and ``FP2" signify the south and north footpoints of the flare loop. The blue and yellow lines in panel (e) represent the positive and negative polarities.
             Intensity contours of the 1600 {\AA} image at 04:57:52 UT are superposed on panel (i) with brown lines.
\label{fig5}}
\end{figure*}

In Figure~\ref{fig5}, nine snapshots of AIA 131 {\AA} images illustrate the converging motion along the hot flare loop. Shortly after the peak times ($\sim$04:58 UT) in UV and radio wavelengths,
the flare loops are empty (see panel (a)). As time goes on, hot plasmas move from the double footpoints (FP1 and FP2) towards the loop top (see panels (c)-(e)). The intensities of the loops gradually 
increase during the upward converging motion and filling process (see panels (g)-(i)). In panel (e), the contours of the positive and negative magnetic polarities at 05:02:08 UT are superposed 
with blue and yellow lines. FP2 and FP1 are rooted in positive and negative polarities, which are associated with the IR and CR (see Figure~\ref{fig3}(d)).

\begin{figure*}
\plotone{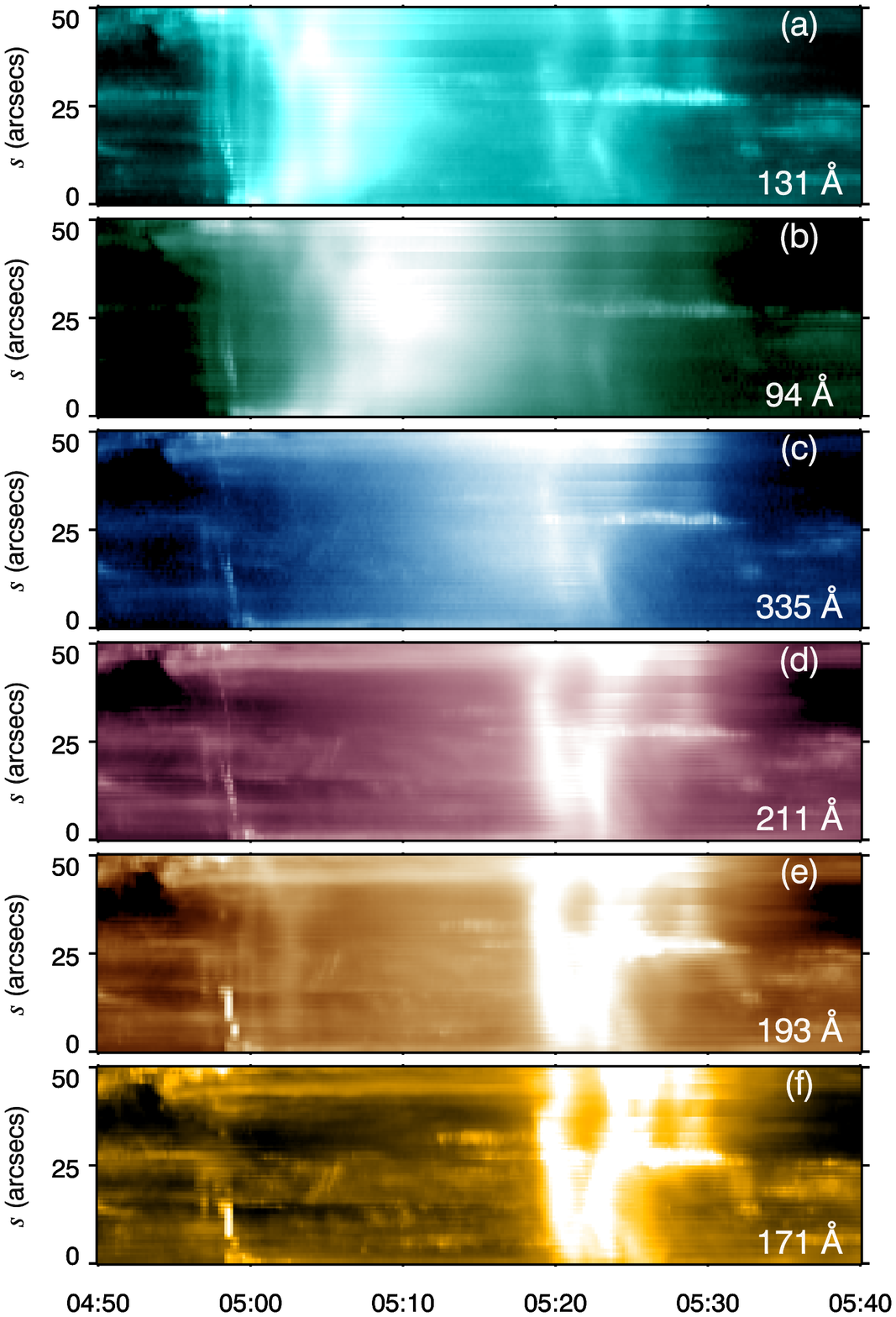}
\caption{Time-slice diagrams of S0 in six EUV wavelengths. $s=0$ and $s=50\arcsec$ in the $y$-axis denote FP1 and FP2, respectively.
            Note that the intensities are in logarithmic scale.
\label{fig6}}
\end{figure*}

To investigate the temporal evolution of the flare loop, we derive the intensities along the curved slice (S0) with a length of 50$\arcsec$ in Figure~\ref{fig5}(a). 
The time-slice diagrams of S0 in six EUV wavelengths are displayed 
in Figure~\ref{fig6}. In panel (a), the converging motion from the footpoints towards the loop top during 04:58$-$05:04 UT is clearly demonstrated. The intensities of flare loops filled with hot plasmas 
reach their maxima around 05:06 UT. Such converging motion and filling process within the flare loops are strongly indicative of chromospheric evaporation in the CRF, which can not been identified 
in the cooler lines.

\begin{figure*}
\plotone{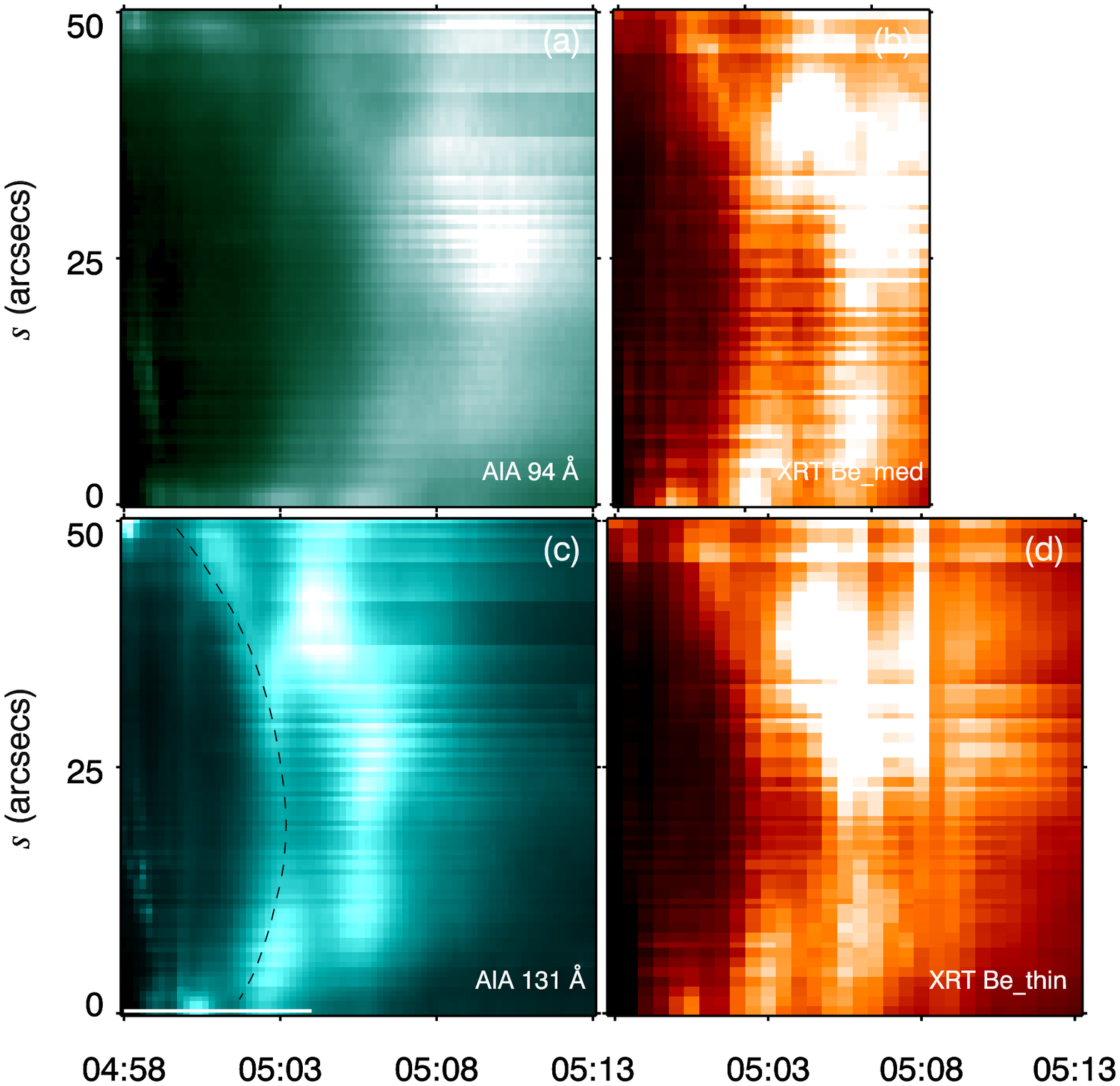}
\caption{Time-slice diagrams of S0 in AIA 94 {\AA} (a), 131 {\AA} (c), XRT Be\_med (b), and Be\_thin (d) filters. In panel (c), the converging motion is outlined by a black dashed line.
          The horizontal line signifies the duration of evaporation from 04:58 UT to 05:04 UT. Note that the intensities are in linear scale.
\label{fig7}}
\end{figure*}

After carefully examining the SXR images observed by \textit{Hinode}/XRT, we found similar filling process of the flare loops. In Figure~\ref{fig7}, the time-slice diagrams of S0 in 94 {\AA} and 
131 {\AA} are displayed in the left panels, and the diagrams in SXR are displayed in the right panels. It is obvious that the converging motion in 131 {\AA} outlined by the black dashed line is 
coincident with that in SXR. In panel (a), there seems to be converging motion in 94 {\AA}. However, we are not quite sure of that since it is very blurring.

\begin{figure}
\plotone{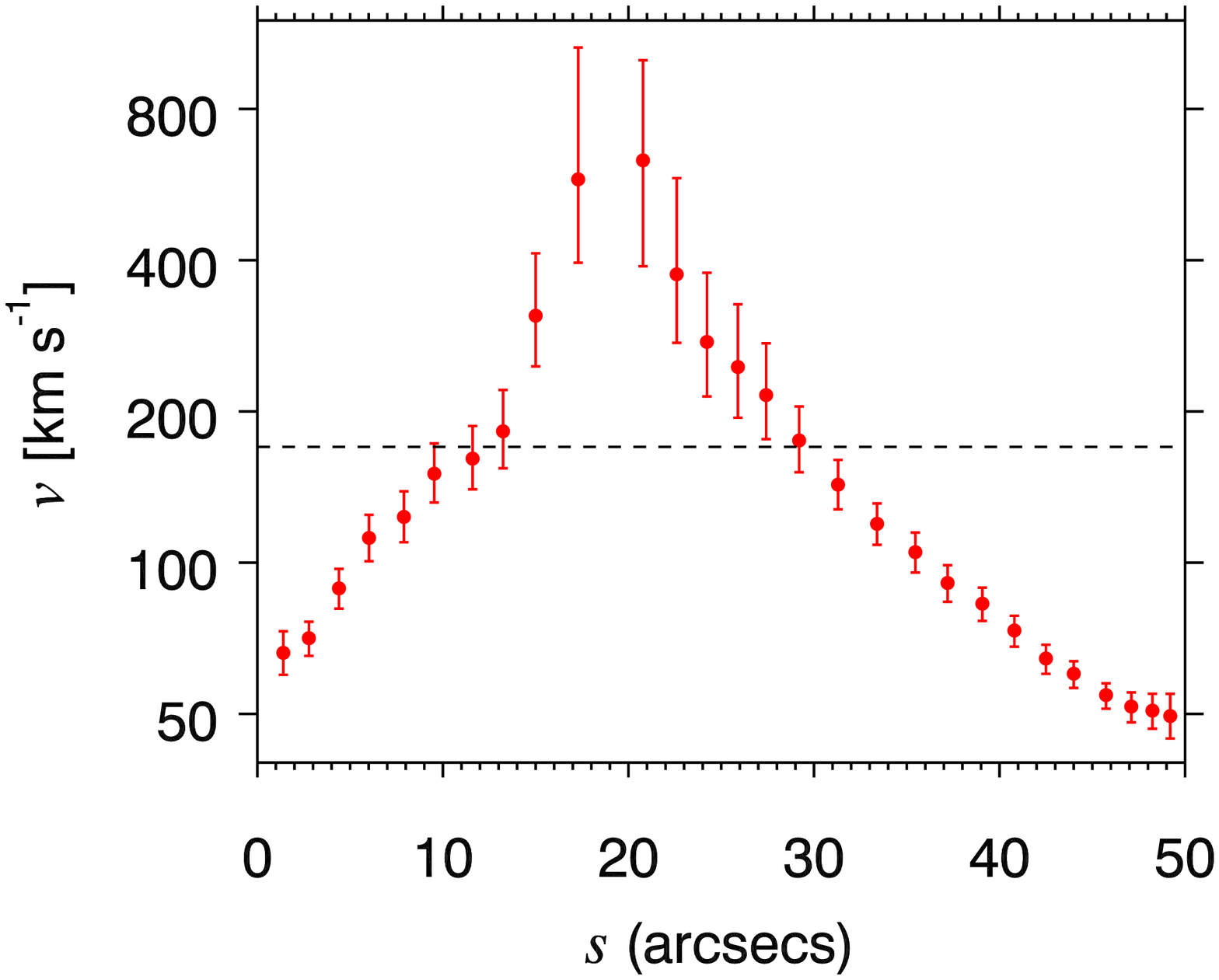}
\caption{Spatial distribution of the apparent velocities of the converging flow along the flare loop. 
The mean value ($\sim$170 km s$^{-1}$) is denoted by the black dashed line.
\label{fig8}}
\end{figure}

The apparent velocities of the converging flow are represented by the slopes ($ds/dt$) of the dashed line in Figure~\ref{fig7}(c). For a certain position $s_i$, 
the velocity is expressed as $v_{i}=(s_{i+1}-s_{i-1})/(t_{i+1}-t_{i-1})$. The uncertainties of velocity come from the uncertainties of time. In Figure~\ref{fig8}, the spatial distribution of the apparent
velocities along the flare loop is plotted with red circles. It is seen that the velocities range from 50 to 630 km s$^{-1}$, with a mean value of $\sim$170 km s$^{-1}$. The error bars of velocity 
increase sharply from the footpoints to the loop apex.

\begin{figure*}
\plotone{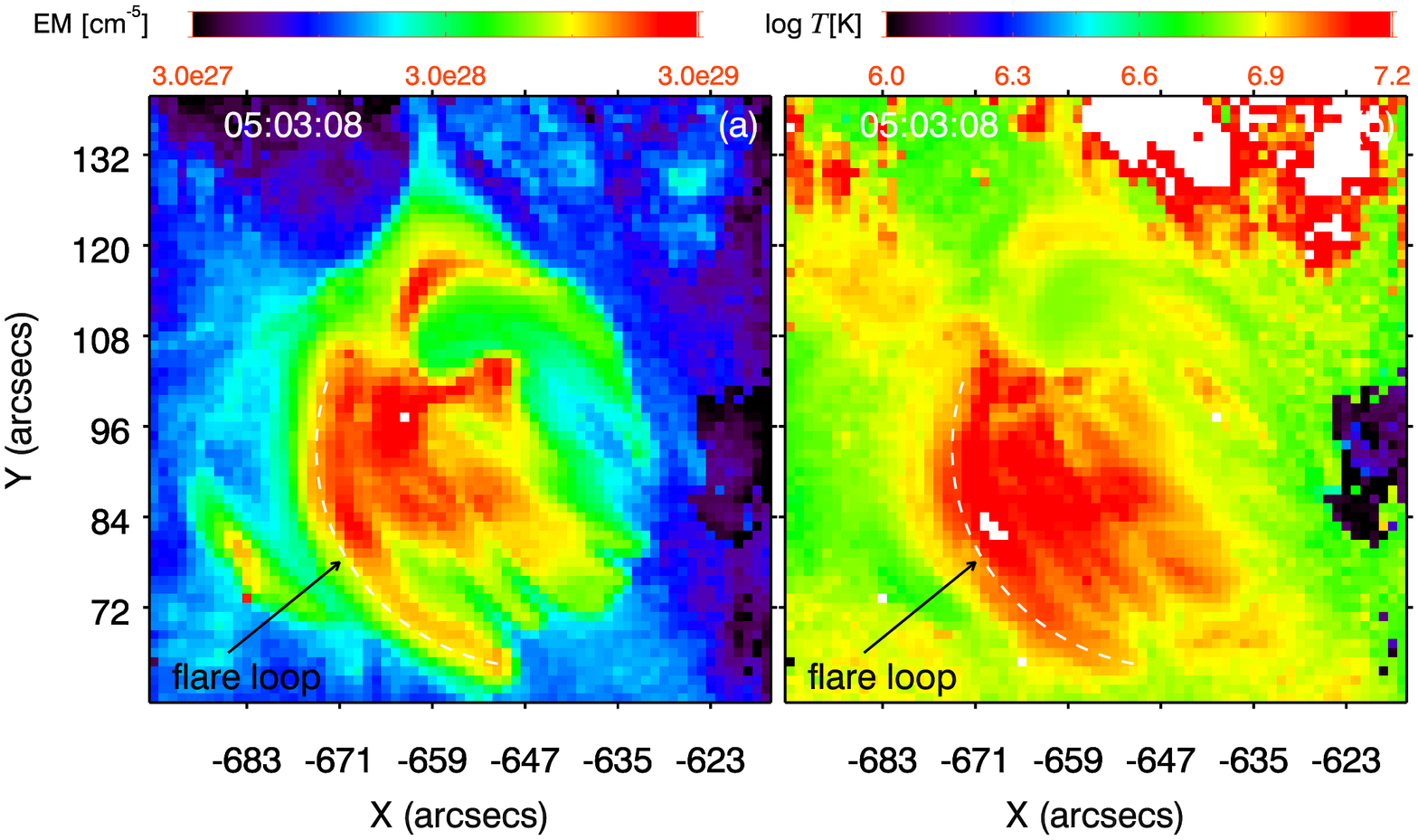}
\caption{Emission measure map and temperature map of the flare at 05:03:08 UT. The black arrows point to the flare loop.
\label{fig9}}
\end{figure*}

Figure~\ref{fig9} shows the emission measure map and temperature map of the flare before the completeness of chromospheric evaporation. The flare region with high density and temperature 
is well reproduced, which is in accordance with the EUV and SXR observations (see Figure~\ref{fig4}). The right panel indicates that the temperatures of flare loops can reach $\sim$10 MK, which 
is in agreement with our evaluation from Figure~\ref{fig6}. It should be emphasized that the flare loop is not a single flux tube, but consist of a bundle of ultrafine strands in deed \citep{jing16}. 
However, the strands could not be precisely distinguished in 131 {\AA}. In Figure~\ref{fig5}(f), we measured the apparent width of the flare loop, which is about 8$\arcsec$. The LOS depth ($H$) equals 
to the width assuming a cylindric flux tube. Taking the value of EM near the loop top ($\sim$1.9$\times$10$^{29}$ cm$^{-5}$), the electron number density ($n_{\mathrm e}$) is estimated to be 
$\sim$1.8$\times$10$^{10}$ cm$^{-3}$ according to Equation~\ref{eqn-2}.

Figure~\ref{fig6} shows that the flare loop became prominent progressively from $\sim$05:06 UT in 131 {\AA} to $\sim$05:20 in 171 {\AA}, which is a clear indication of 
cooling process. Considering that thermal conductive cooling dominates over radiative cooling for hot plasmas ($\sim$10 MK), the cooling timescale is expressed as
\begin{equation} \label{eqn-4}
\tau_c=4\times10^{-10}\frac{n_{\mathrm{e}}L^2}{T_{\mathrm{e}}^{5/2}},
\end{equation}
where $n_\mathrm{e}$, $T_\mathrm{e}$, and $L$ represent the electron number density, temperature, and total length of a coronal loop \citep{car94}. 
According to the estimated values of $n_\mathrm{e}=1.8\times10^{10}$ cm$^{-3}$, $T_\mathrm{e}=10^7$ K, and $L=5.7\times10^9$ cm with a semicircular shape, 
$\tau_c$ is estimated to be 12 minutes, which is roughly consistent with the cooling time (14 minutes) of the flare loop.

\section{Discussion} \label{s:disc}
Chromospheric evaporation has been extensively studied in the past three decades. The temperatures of evaporation upflows can reach tens of millions degrees \citep[e.g.,][]{you13,tian14,li15,pol15,zqm16a}.
In this study, the converging motion from the footpoints to the loop top and filling process in the flare loop are simultaneously observed by \textit{SDO}/AIA in 131 {\AA} and the SXR filters on board XRT.
The hot evaporation upflow is further justified by the DEM analysis. During the impulsive phase of C4.2 CRF on 2015 October 16, upflows at speeds of 35$-$120 km s$^{-1}$ on the flare ribbons are 
detected in the Fe {\sc xxi} $\lambda$1354.09 emission line ($\log T\approx7.05$) \citep{zqm16a}. Hence, the spectroscopic and imaging observations consolidate the existence of high-temperature
evaporation upflows in CRFs. Using the apparent width of the flare loop, we also estimated the electron number density of the evaporation upflow. The value (1.8$\times$10$^{10}$ cm$^{-3}$)
is 1$-$2 orders of magnitude lower than that in the flare loops on 2014 October 27 \citep{pol16}. One plausible reason is that the estimated value is a lower limit assuming that the filling factor of 
plasma equals to 1.0.

In previous studies of direct imaging observations of chromospheric evaporation in EUV and SXR wavelengths, the velocities of upflows range from 100 to 500 km s$^{-1}$ \citep{sil97,nit12,li17a}.
Another way of detecting the chromospheric evaporation is to track the HXR footpoint sources along the flare loops. The velocities of drifting or converging motions of the sources are reported to be 
a few hundred km s$^{-1}$ \citep{liu06,ning09,ning10}. In this study, the flare loops are observed head-on rather than edge-on. Therefore, the measured velocities (50$-$630 km s$^{-1}$)
of the evaporation upflows are projected or apparent velocities. The true values, after correcting the projection effect with the assumption of a semicircular shape, should be larger by a factor of 
$\sim$1.5 near the footpoints. Since spectroscopic observations of the flare focused on the eastern edge of CR with weak intensities \citep{zqm18}, precise Doppler velocities of the upflows could 
not been obtained. Anyway, the velocities of the upflows are in accordance with previous findings.

As to the driving mechanism of chromospheric evaporation, the roles of nonthermal electrons and thermal conduction have been largely investigated \citep[e.g.,][]{nag84,abb99,all05,bat09,reep15}. 
During the C4.2 CRF as mentioned above, explosive chromospheric evaporation occurred on both CR and IR \citep{zqm16a}. Based on the quantitative calculation of electron energy flux 
and the spatial correspondence between the HXR source and IR, the authors concluded that the evaporation was driven by nonthermal electrons accelerated by magnetic reconnection. 
The estimated electron energy flux ((1$-$4)$\times$10$^{10}$ erg cm$^{-2}$ s$^{-1}$) is sufficient to drive explosive evaporation as predicted by theory. 

For the C5.5 flare in our study, direct HXR observations were unavailable during the impulsive phase. However, chromospheric evaporation occurs shortly after the coincident peak times 
in 1600 {\AA}, 17 GHz, and SXR derivative, implying that the chromosphere responds very quickly to the impulsive energy deposition \citep{zqm16b,kum16,hao17,song18}. 
Using one-dimensional hydrodynamic numerical simulations, \citet{reep15} investigated the importance of electron energy on the explosive and gentle evaporations and on the atmospheric response. 
It is found that for explosive evaporation, the atmospheric response does not depend strongly on electron energy, while for gentle evaporation, lower energy electrons are more efficient at heating the 
atmosphere and driving upflows. Comparing their results of maximal density, temperature, and velocity of the upflows across a broad range of electron energy (5$-$50 keV) and a broad range of energy 
flux (10$^8$$-$10$^{11}$ erg cm$^{-2}$ s$^{-1}$), it is inferred that the electron energy flux above 5$\times$10$^{10}$ erg cm$^{-2}$ s$^{-1}$ are likely needed to explain our observations.
It should be emphasized that the comparison between observations and simulations (even though this is beyond the scope of the present work) requires a proper forward modeling of different 
observables.

In addition to the total energy flux, the duration of heating is an important parameter that has effect on the atmospheric response. \citet{reep18} investigated the role of electron heating duration.
It is found that the duration of upflows act as a good diagnostic of heating duration. In a multithreaded model, durations of 100$-$200 s can well reproduce both the red- and blueshifts for a fixed 
heating \citep[see also][]{war06}. In Figure~\ref{fig1}(b), the heating duration of $\sim$240 s is labeled with an orange arrow, which is derived from the full width at half maximum (FWHM) 
of the HXR proxy. The heating duration is close to the results of numerical simulations. In Figure~\ref{fig7}(c), the time of converging motion from 04:58 UT to 05:04 UT is represented by a 
horizontal line. The decay time of upflow seems to be longer than the heating duration, which is probably due to the larger energy flux.

\citet{reep16} developed a numerical model of flare heating due to the dissipation of Alfv\'{e}nic waves propagating from the corona to the chromosphere. The waves damp themselves 
while propagating along the flux tubes as a result of collisions between electrons, ions, and neutrals, which decrease the wave amplitude and heat the local plasma. It is found that waves with 
sufficiently high frequencies and perpendicular wave numbers are able to heat the upper chromosphere and the corona. The temperatures of upper chromosphere and corona can rise up to 
$\sim$10$^5$ K and a few MK within 10 s. Meanwhile, the heating can drive explosive evaporation, with the maximal blueshifted and redshifted velocities being $\sim$200 km s$^{-1}$
and $\sim$20 km s$^{-1}$. Hence, the atmospheric response to Alfv\'{e}nic wave heating is similar to that of heating by electron beam with a low-energy cutoff of 20 keV, a spectral index of 5, 
and an energy flux of 10$^{10}$ erg cm$^{-2}$ s$^{-1}$ (see their Fig. 1). Without HXR imaging observations, it is impossible to distinguish the contributions of nonthermal electron and Alfv\'{e}nic waves.

The energy flux of thermal conduction is expressed as
\begin{equation} \label{eqn-5}
F_c\approx \kappa_0 T_{\mathrm e}^{7/2}/L,
\end{equation}
where $\kappa_0\approx$10$^{-6}$ erg K$^{-2/7}$ cm$^{-1}$ s$^{-1}$ and $L$ is the flare loop length \citep{wue94}. 
For the flare loop of CRF, $F_c$ is estimated to be 5.5$\times$10$^8$ erg cm$^{-2}$ s$^{-1}$, which is much lower than the requirement of electron beam energy flux. 
The timescale of thermal conduction is measured by the propagation time of thermal conduction front from the loop apex to the footpoints. 
The thermal conduction velocity is expressed as
\begin{equation} \label{eqn-6}
v_0=\frac{2\kappa_0}{3k_B}\frac{T_{\mathrm e}^{5/2}}{n_\mathrm{e}L_0},
\end{equation}
where $k_B$ is the Boltzmann constant, $L_0$ represents the temperature scale height \citep{rust85}. Assuming that $L_{0}=6000$ km, $v_0$ is estimated to be $\sim$1400 km s$^{-1}$. 
The timescale of thermal conduction is $\sim$40 s. Therefore, thermal conduction as the main mechanism to heat the flare can be ruled out.

\section{Summary} \label{s:sum}
In this work, we report our multiwavelength observations of the C5.5 CRF on 2014 August 24. The main results are summerized as follows:
\begin{enumerate}
\item{The CRF was related to two coronal jets (jet1 and jet2) that propagated along large-scale closed loops. Jet1 appeared first with untwisting motion, while jet2 was generated 6 minutes later.
The CRF consisted of a discrete CR with a diameter of $\sim$1$\arcmin$ and a short IR inside. They were observed in UV, EUV, SXR, and especially in 17 GHz. 
The bright patches in 17 GHz were cospatial with the flare ribbons. The CR and IR were associated with negative and positive polarities, implying a magnetic null point in the corona.}
\item{The peak times of the flare in 1600 {\AA}, 17 GHz, and SXR derivative were coincident at $\sim$04:58 UT, indicating the peak time of impulsive energy deposition in the lower atmosphere.
Converging motion from the footpoints towards the loop top and filling process in the flare loops are revealed in AIA 131 {\AA} and XRT filters (Be\_thin and Be\_med), which are clear evidence for 
chromospheric evaporation. The upflows started from $\sim$04:58 UT until $\sim$05:04 UT, covering the SXR peak time ($\sim$05:02 UT). 
The temperature, density, and apparent velocities of upflows are $\sim$10$^7$ K, 1.8$\times$10$^{10}$ cm$^{-3}$, and 50$-$630 km s$^{-1}$ with a mean value of 170 km s$^{-1}$.
The flare loops cooled down via thermal conduction with a timescale of 14 minutes and appeared progressively from 131 {\AA} to 171 {\AA}.}
\item{Despite of the lack of HXR observations, the requirement of electron energy flux is estimated to be above 5$\times$10$^{10}$ erg cm$^{-2}$ s$^{-1}$, while the energy flux of thermal 
conduction in heating the chromosphere is estimated to be 5.5$\times$10$^8$ erg cm$^{-2}$ s$^{-1}$. Since heating by thermal conduction does not seem to provide enough energy,
alternative mechanisms such as nonthermal electrons or Alfv\'{e}nic waves might need to be invoked.}
\end{enumerate}

\begin{acknowledgements}
The authors appreciate the referee for valuable comments and suggestions to improve the quality of this article.
We would also like to thank A. Warmuth and Z. J. Ning for fruitful discussions.
\textit{SDO} is a mission of NASA\rq{}s Living With a Star Program. AIA and HMI data are courtesy of the NASA/\textit{SDO} science teams. 
This work is funded by NSFC (Nos. 11333009, 11790302, 11773079, 11603077, 11203083, U1731241, 11729301), the Fund of Jiangsu Province (BK20161618, BK20161095), 
and the Strategic Pioneer Program on Space Science of CAS (XDA15052200 and XDA15320301).
Q.M.Z is supported by the Youth Innovation Promotion Association CAS. D.L. is supported by the Specialized Research Fund for State Key Laboratories.
\end{acknowledgements}

\end{document}